\documentclass[a4paper,10pt]{article}

\usepackage{amssymb,latexsym,amsmath,amsfonts}
\usepackage{graphicx}
\usepackage{subfigure}
\usepackage{color}
\usepackage[english]{babel}
\usepackage{authblk}
\usepackage{subfigure}
\usepackage[utf8]{inputenc}
\usepackage{pdfpages}
  \newcommand{\pardt}{\partial t}
\newcommand{\pardx}{\partial x}

\newcommand{\pardy}{\partial y}

\newcommand{\be}{\begin{equation}}
\newcommand{\ee}{\end{equation}}

\title{Viscous-Inviscid Interactions in a Boundary-Layer Flow Induced by a Vortex Array}
\author[D. Kopta et al.]
       {F. Gargano$^{1,2}$, M. Sammartino$^1$, V. Sciacca$^{1}$,
       K.W~Cassel$^3$
       \\
      \tiny{ $^1$National Research Council (CNR), Institute for Coastal Marine Environment (IAMC),\\
             Via L. Vaccara 61, 91026, Mazara del Vallo (TP), Italy.\\
       $^2$Dept of Mathematics and Computer Science,  University of Palermo\\ Via Archirafi 34, 90123
 Palermo, Italy.\\
       $^3$Department of Mechanical, Materials, and Aerospace Engineering, \\ Illinois Institute of Technology,\\
            10 West 32nd Street, Chicago, Illinois, USA.\\
       }
       
       \small{\textit{Email addresses:}\\
       francesco.gargano@unipa.it (F.G),\\
       marco@math.unipa.it (M.S.),\\
       sciacca@math.unipa.it (V.S),\\
       cassel@iit.edu (K.C)}}

\begin{document}
\date{}
\maketitle


\begin{abstract}
In this paper we  investigate 
the asymptotic validity of boundary layer theory. For a flow induced by a periodic row of point-vortices, we compare Prandtl's solution to Navier-Stokes solutions at different
$Re$ numbers. We show how Prandtl's solution develops a finite time separation singularity.
On the other hand Navier-Stokes solution is characterized by the presence of two kinds of viscous-inviscid interactions between the boundary layer and the outer flow.
These interactions  can be detected
by the analysis of the enstrophy and of the pressure gradient on the wall.
Moreover we apply the complex singularity tracking method to  Prandtl and Navier-Stokes 
solutions and analyze the previous interactions from a different perspective.
\end{abstract}

\section{Introduction}

The study of the behavior of a high Reynolds number ($Re$) fluid interacting with a
solid boundary is a central problem
in the mathematical analysis of fluid dynamics. Due to the no-slip boundary condition, 
the convergence of Navier-Stokes (NS) solution to Euler solution, as $Re\rightarrow\infty$, fails, and Prandtl boundary layer equations must be used
to resolve the flow close to the boundary.
We mention the papers \cite{SC98a,SC98b,CS97,LCS01}  where, for analytic 
initial data, the authors prove the convergence  of NS solutions to
Euler and Prandtl. The zero viscosity limit was also  considered in \cite{Kato84,TW97,Kel07}, where the
authors introduce criteria based on a priori estimates of energy dissipation in a viscous sub layer, and in \cite{Mae13} where the author employs the assumption that,
initially, the vorticity close to the
boundary is zero. Strong convergence of Navier-Stokes to Euler solutions in $L^2$ 
spaces is given in \cite{LML08,LMLT08}  with a symmetry assumption. 
In \cite{CMR98,LLP05,Kel06,IP06,WXZ13,BC11,BC12,MS12,CKV14},  
the authors study the inviscid limit of the two-dimensional Navier-Stokes equations in the case of a Navier friction 
boundary condition.
Conditions on the well-posedness of Prandtl's equation
are given in \cite{CLS01,LCS03,XZ04,KV13,CLS13}, while results on the ill-posedness are in \cite{GVD10,GVN12}. The problem of the well posedness of Prandtl's equation
is also related to the finite time singularity formation that has been observed for many significant flows (see for example  \cite{vDS80,PSW91a,Cas00}).  
At the singularity time  the normal component of the
velocity   becomes infinite  with  ejection of vorticity and flow
particles from within the boundary layer into the outer flow;  the consequent
breakdown of the assumptions on which Prandtl's equation are based on signals the limit of the classical boundary layer theory.
However 
Prandtl and NS solutions begin to show a quantitative disagreement prior to Prandtl singularity formation: in fact, various interactions between 
the viscous boundary layer and the inviscid outer flow develop in NS flow.
These interactions  behave in a different manner from that observed in the Prandtl boundary-layer, and
as observed in other initial flows (see \cite{Cas00,GSS11,GSSC14}), they act over different length scales and they influence the flow evolution in a very different manner.
The first interaction, called  large-scale ($LS$) interaction, is found to occur for
all finite Reynolds numbers, and it signals the time when the comparison between Navier-Stokes and Prandtl's solutions
begins to show some quantitative discrepancies.
The  small-scale ($SS$) interaction  develops only for moderate to high Reynolds numbers 
(generally $Re\geq O(10^4)$). This interaction  
is marked by large gradients along the streamwise to the boundary variable,
a chaotic formation of small-scale vortical structures and a large amount of vorticity production on the boundary that, in turn, leads to the growth of enstrophy.
This growth, caused by the collision on the wall of the various vortical structures
that forms during the separation process, is absent both in Prandtl solutions
as well in NS solutions for low $Re$.\\
In this paper we shall investigate these interactions in  a flow induced by
a vortex array.
\section{{Statement of the problem and numerical procedures}}
\label{sc_initialdatum}

Our case study  consists
of an infinite row of point-vortices immersed in a 2D viscous incompressible flow
at rest at infinity and bounded by an infinite rectilinear wall.
In a cartesian frame the vortices are centered in
($ma+\pi$, $b$)$_{m \epsilon \mathbb{Z}}$, where $b$ is the
distance of the  row from the wall and $a$ is the distance of
between two consecutive vortices. All the vortices have strength $k$. 
We study the system in the reference frame comoving  with the vortices. The initial data for the streamwise and normal velocity components are 
$u_0=\partial_y \Psi_E, v_0=-\partial_x \Psi_E$,
where 
\begin{equation}
   \Psi_E(x,y)=-U_c y-\frac{k}{4\pi}\log(\frac{\cosh(\frac{2\pi}{a}(y-b/2))-\cos(\frac{2\pi}{a}(x-\pi))}
   {\cosh(\frac{2\pi}{a}(y+b/2))-\cos(\frac{2\pi}{a}(x-\pi))}) \label{streameqgargh}
\end{equation}
is the  streamfunction of the inviscid steady Euler solution for the vorticity configuration described above, while  $U_c=\frac{k}{2a}\cosh(\pi b/a)$, see  \cite{Lamb}.
This is an $a$-periodic datum, and the velocity components
obtained are such that $u =k/a$, $v = 0$ at $y=0$, 
$u, v \rightarrow 0$ for $y \rightarrow \infty$.
We set $a=2\pi,k=2a$ and we solve our problem in the domain
[0,2$\pi$]x[0,$\infty$).
Using $b$ and
$U_c$  as characteristic length and velocity,
 NS equations in the vorticity-streamfunction formulation are:
\begin{eqnarray}
\frac{\partial \omega}{\pardt}+u\frac{\partial \omega}{\pardx}+v\frac{\partial \omega}{\pardy}&=&
\frac{1}{Re}\Delta \omega, \label{NSequation}\\
\Delta \psi&=&-\omega, \label{poisson}\\
u=\frac{\partial \psi}{\pardy}, \quad v&=&-\frac{\partial \psi}{\pardx},\label{velocityew}\\
\omega(x,y,t=0)=\omega_0&=&4\pi\delta_{(\pi,1)},\quad \psi(x,y,0)=\Psi_E \label{NSinit}\\
\omega(x,y\rightarrow\infty,0)&=&0 \label{vorticitybc}, \quad \omega(0,y,t)=\omega(2\pi,y,t)\\
u(x,0,t)=-\coth (1),\quad v(x,0,t)&=&0 . \label{noslipppp}
\end{eqnarray}
In the above equations the  Reynolds number  is defined as $Re=bU_{c}/\nu$, being $\nu$ the kinematic
viscosity.

The initial vorticity is singular and to avoid this initial singularity
we approximate the initial solution by convolving it with the gaussian mollifier
 $\Phi^{\sigma}(x,y)
=\frac{1}{\sigma^{2}\pi} e^{- \frac{(x^2 + y^2)}{\sigma^2}}$,
obtaining the regularized vorticity
$
\omega^{\sigma}=\omega_0\ast\Phi^{\sigma}=4\pi \Phi^{\sigma}(x-\pi,y-1). \label{vorticityinitgargh}
$
This is a typical procedure used in computational vortex methods when the initial
data have point singularities (as in \cite{GSS11}) or for vortex-sheet
motion (as in \cite{Sh92}). In the NS calculations we have
chosen $\sigma^{2}=5\times 10^{-3}$. As $\psi \rightarrow \infty$ for $y\rightarrow \infty $
we truncate the computational domain  at a finite
value $y_{\max}$ of the normal variable. Following what proposed in \cite{GSS11} for the rectilinear vortex case,
we chose this value requiring that the
vorticity remains negligible for $y\geqslant y_{\max}$ for all
computational time.
We take $y_{\max} = 6$ for all the $Re$ numbers, which we find to be large enough
to satisfy the required condition for the vorticity as we have checked that for $y>6$ no relevant differences arise between the various solutions. 

Prandtl's equations for this setup are:
\begin{eqnarray}
\frac{\partial u}{\partial t}+u\frac{\partial u}{\partial x}+V
\frac{\partial u}{\partial Y}-U_{\infty}\frac{\partial
U_{\infty}}{\partial x}&=&\frac{\partial^2 u}{\partial Y^2}, \label{pramomentum}\\
\frac{\partial u}{\partial x}+\frac{\partial V}{\partial Y}&=&0,\label{consmass}
u(x,Y,0)=U_{\infty},\label{initialprandtl}\\
u(x,0,t)&=&-\coth (1), \quad u(x,Y\rightarrow\infty,t)=U_{\infty},\label{boundaryprandtl}\\
u(0,y,t)&=&u(2\pi,y,t),\label{praperiod}
\end{eqnarray}
where $U_\infty=\partial_y \Psi_{E|y=0}$ is the inviscid solution of the outer flow.\\ 

The system \eqref{NSequation}-\eqref{noslipppp} is solved 
with a Galerkin-Fourier method 
in the streamwise variable, while the Chebyshev-collocation method is used in the normal
variable.  This ensures fully spectral convergence, see \cite{Pey}. The temporal
discretization used is the  Adams-Bashforth-Implicit Backward Differentiation method, and to find 
the necessary vorticity boundary condition, 
the influence matrix method \cite{Pey} is used. 
The description of the numerical  techniques used to treat the initial singular vorticity and the infinite normal domain
can be found in \cite{GSS11}. Numerical solutions for
Reynolds numbers ranging from $10^3$ up to $5\cdot10^4$ are computed, with computational
grids up to $8192\times1025$ points for the higher Reynolds numbers.
To solve Prandtl system  \eqref{pramomentum}-\eqref{praperiod} we have used the fully spectral numerical scheme
 used in \cite{GSS09} and a computational grids up to $4096\times1025$ points. 
 
\section{Prandtl's Solution}
In this section we shall describe the main physical events leading to singularity formation for Prandtl's equation, and 
we shall apply the singularity tracking method to detect the time at which singularity forms.

\begin{figure}
\begin{center}
\subfigure[$t=0.5$]{\hskip-0.35cm\includegraphics[width=8cm]{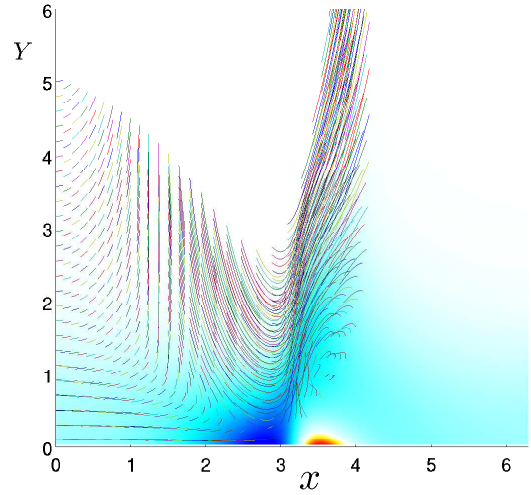}\label{prarec}}
\subfigure[$t=1$]{\hskip-0.35cm\includegraphics[width=8cm]{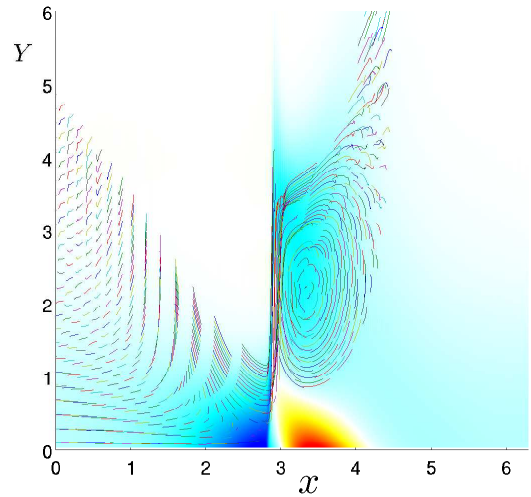}\label{prasing}}
\caption{The vorticity contour levels of Prandtl's solution at time $t=0.5$ and time $t=1$ (red colors positive vorticity, blue color negative vorticity).
In these figures the pathlines of flow particles are also show. In $a)$ the recirculation region has already formed, and the pathlines followed from $t=0.4$ to $t=0.5$,
begin to rotate clockwise in the region above the positive vorticity.  In $b)$, just prior the singularity formation, a spike is visible in the vorticity contour,
and the pathlines, followed from $t=0.9$ to $t=1$ experience a rapid transition in the normal direction in the streamwise position $x_{s}\approx2.86$}
\end{center}
\end{figure}

The formation of a recirculation eddy detached from the wall is the first relevant physical event in Prandtl's solution. It occurs at $t\approx0.28$,
and it is due to the adverse streamwise pressure gradient (the term $U_{\infty}\frac{\partial
U_{\infty}}{\partial x}$ in \eqref{pramomentum}) imposed by the outer flow . This recirculation region can be observed in Fig.\ref{prarec} where the 
Prandtl's  vorticity $\omega_P=-\partial_Y u$ is shown  at $t=0.5$, along with some pathlines of fluid particles followed from $t=0.4$ up to $t=0.5$. The
pathlines  rotate clockwise above the zone of positive vorticity where the recirculation region is formed.
As time passes this eddy thickens rapidly in
the streamwise direction, and a  spike in vorticity contours forms at $t\approx0.86$. 
This is visible in Fig.\ref{prasing} at $t=1$
where also the  pathlines are shown: the pathlines experience 
a rapid transition in the normal direction, meaning that fluid particles
are ejected from within the boundary layer to the outer flow.

To characterize the singularity of Prandtl's solution, we apply the singularity tracking method
 (see \cite{SSF83}). This method 
has been widely used to characterize the singularity formation for ODEs and PDEs (see \cite{SSF83,Wei03,PMFB06,DLSS06,GSS09,CGS12}).
We  write the Fourier-Chebyshev expansion of Prandtl's solution (see \cite{GSS09} for the details): 
\begin{equation}
u(x,y,t)\approx \sum\limits_{k=-M/2}^{k=M/2}\sum\limits_{j=0}^{j=N}u_{kj}(t)e^{ik\theta}T_j(y)\; ,
\label{ChebT}
\end{equation}
where $T_j(y)$ are Chebyshev polynomials of the first kind. In this way we can apply the singularity tracking method
to the shell-summed Fourier amplitudes defined as
\begin{equation}
A_K \equiv \sum_{K\leq |(k,j)| < K+1} |u_{k,j}| \label{sing_shellSUMampl},
\end{equation}
which have the following asymptotic behavior (\cite{PMFB06}): 
\begin{equation}
A_K\approx CK^{-\left(\alpha_P+1\right)} \exp{\left(-\delta_P K\right)}\label{shella},\quad \text{for } K\rightarrow \infty.
\end{equation}
In the above formula $\delta_P$ gives the width of the analyticity strip while the algebraic
prefactor $\alpha_P$ gives informations on the nature
of the singularity. By performing a fitting procedure on the parameters $\delta_P$ and $\alpha_P$, we find
that  $\delta_P\approx0$ at $t_s\approx 1$, revealing the singularity formation,
while the characterization of the singularity at $t_s$ is $\alpha_P\approx1/3$.
We notice that  the characterization $\alpha_P\approx1/3$  is the same obtained in \cite{GSS09} for the impulsively started disk. 
\vspace*{-0.25cm}\section{Navier--Stokes Solution}
In this section we compare Navier-Stokes solutions at different
$Re$ numbers ($10^{3},10^{4}$ and $5\cdot10{^4}$), with the solution
of the boundary layer equations. Moreover we shall describe the viscous--inviscid
interactions between the viscous boundary layer and the outer flow which characterize 
Navier-Stokes solutions.
\subsection{LS interaction}
\begin{figure}
\begin{center}
\subfigure[$Re=10^3$]{\hskip-0.3cm\includegraphics[width=8.05cm]{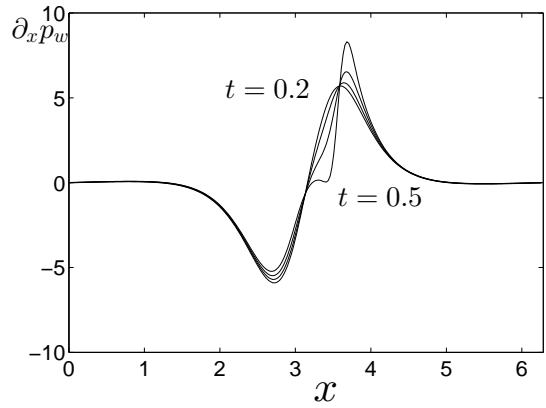}\label{dpdx103}}
\subfigure[$Re=5\cdot10^4$]{\includegraphics[width=8.05cm]{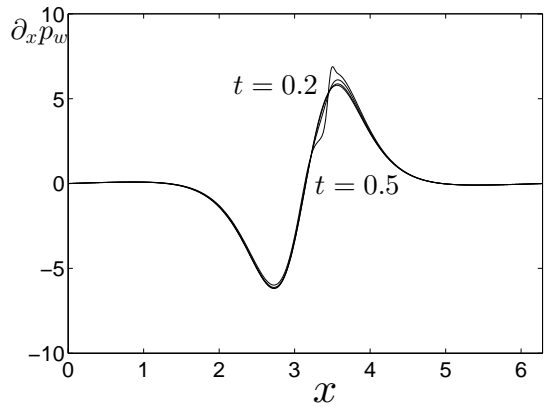}\label{dpdx105}}
\caption{ The time evolution of $\partial_{x}p_{w}$
for  $Re=10^3,5\cdot10^4$ number (time step of 0.1). Large--scale interaction begins when an inflection point forms in $\partial_{x}p_{w}$, 
and this happens at time $t_{LS}\approx0.33,0.38$ 
for $Re=10^3,5\cdot10^4$ respectively.}
\end{center}
\end{figure}

\begin{figure}
\begin{center}
\subfigure[$Re=10^3,t=0.5$]{\hskip-0.0cm\includegraphics[width=6cm]{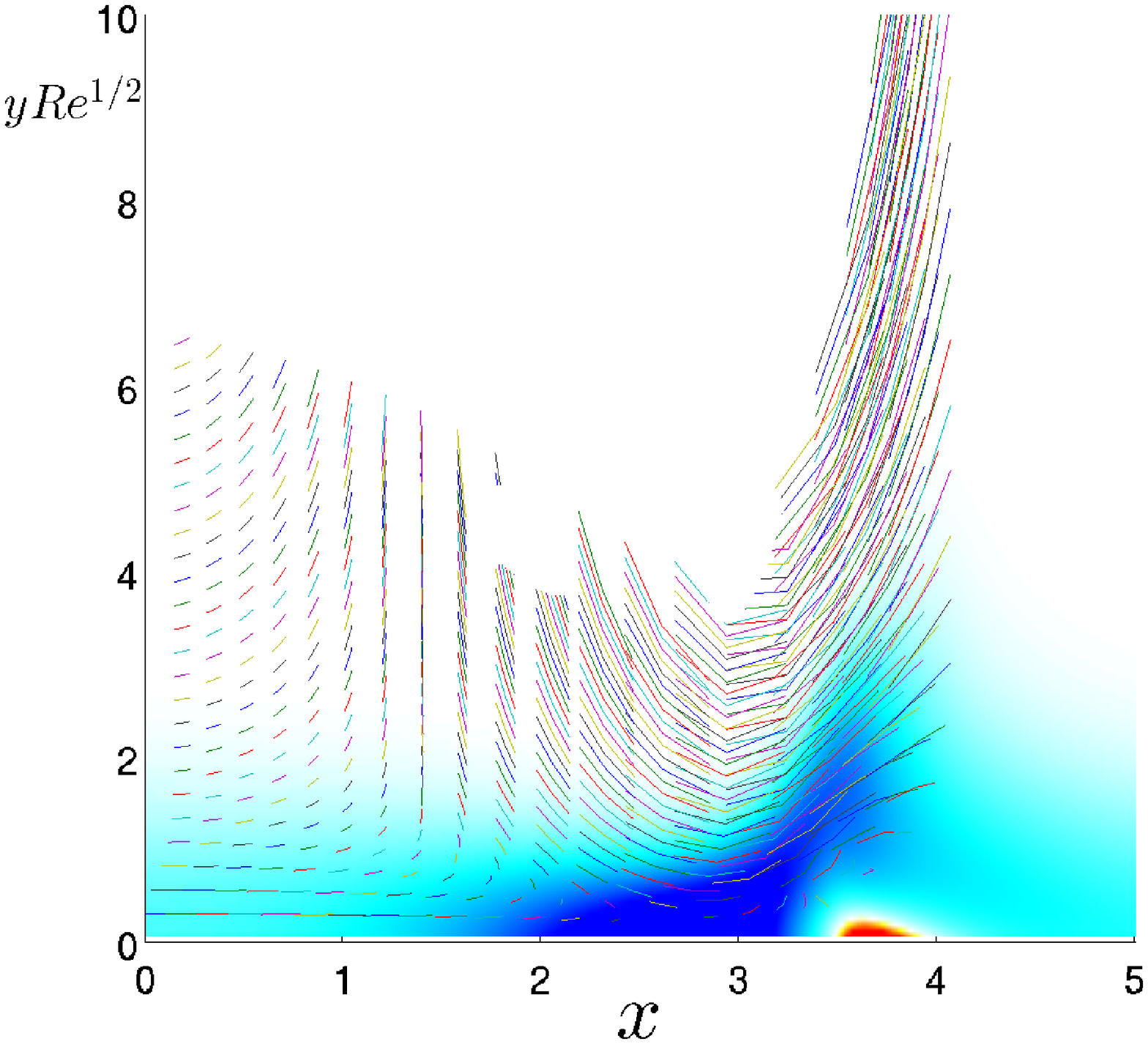}\label{vor103_t05}}
\subfigure[$Re=10^3,t=1$]{\hskip-0.3cm\includegraphics[width=6cm]{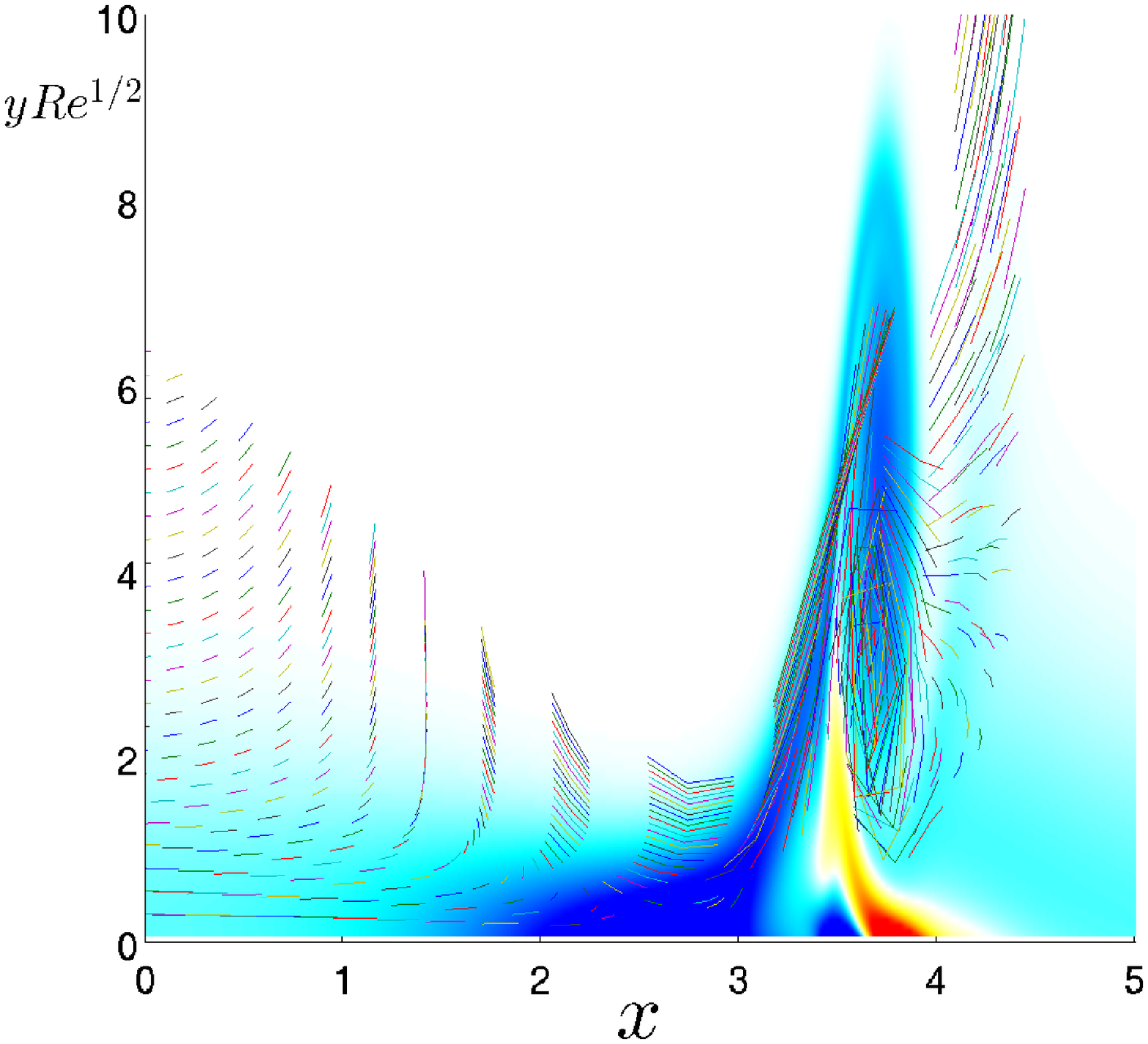}\label{vor103_t1}}
\subfigure[$Re=10^3,t=2$]{\hskip-0.0cm\includegraphics[width=6cm]{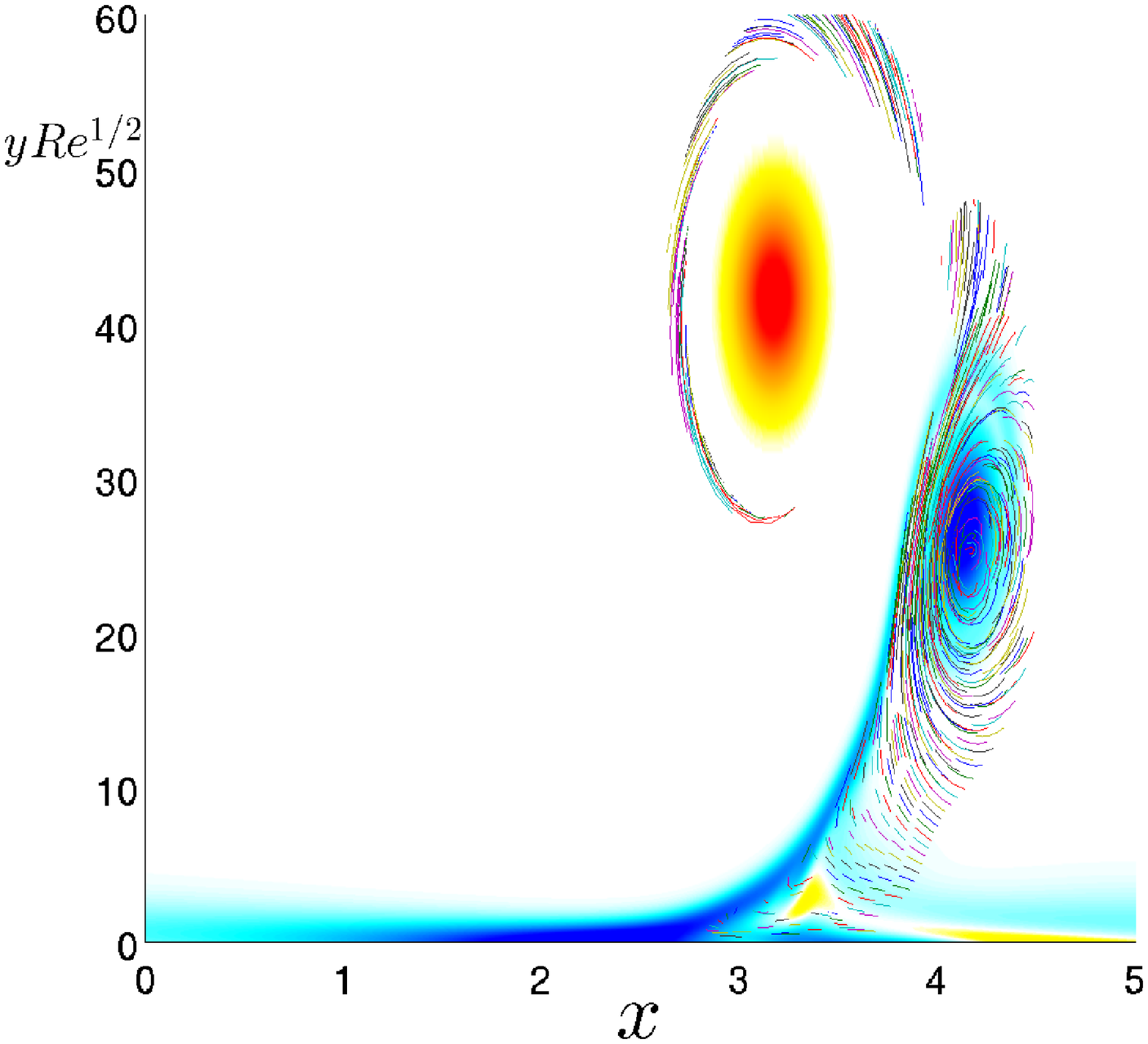}\label{vor103_t2}}
\subfigure[$Re=10^3,t=3$]{\hskip-0.3cm\includegraphics[width=6cm]{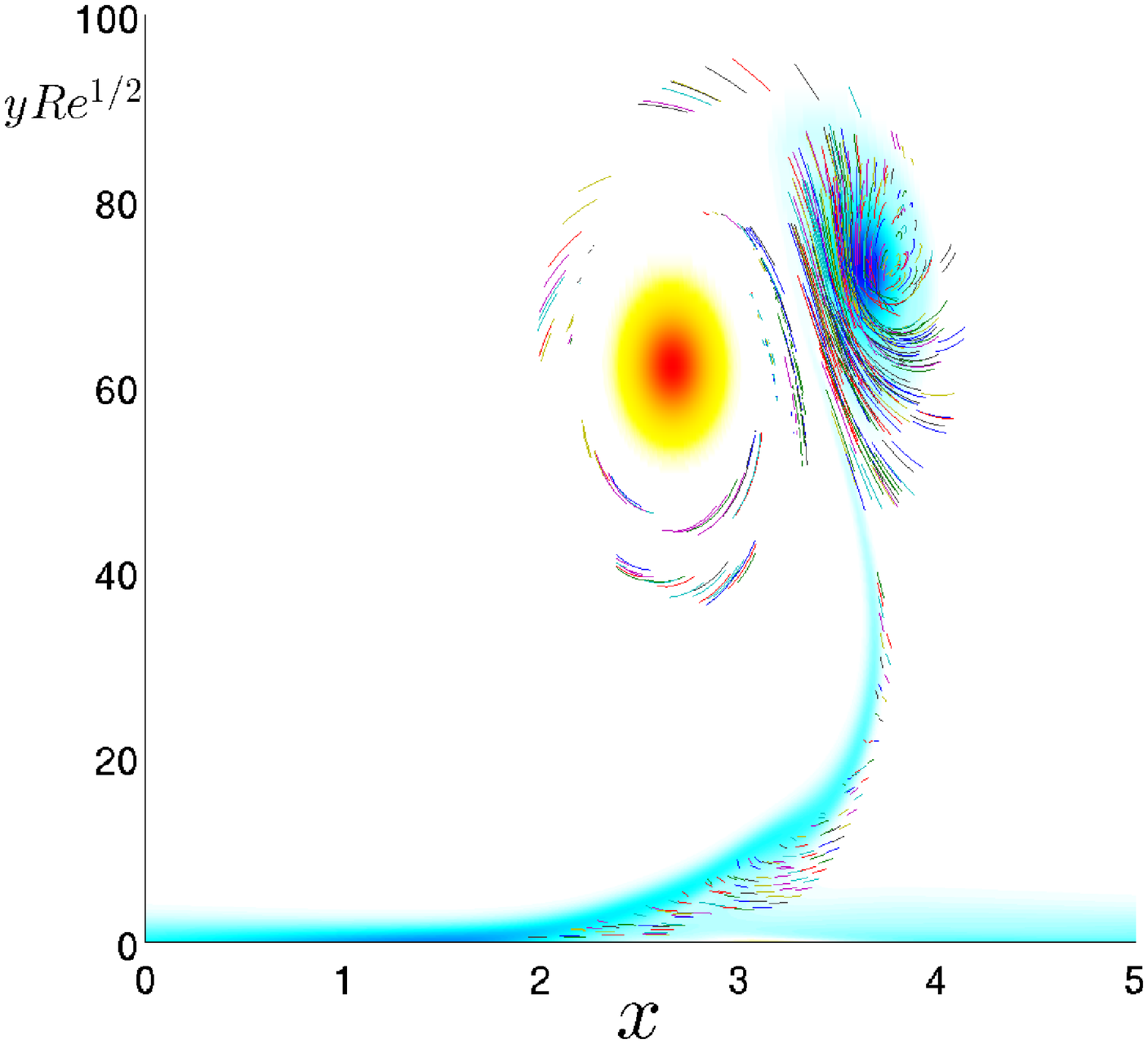}\label{vor103_t3}}
\caption{ The vorticity (red colors positive vorticity, blue color negative vorticity) and the pathlines of fluid particles  for $Re=10^3$
at time $t=0.5,1,2,3$. The pathlines are followed in temporal intervals $[0.4,0.5],[0.9,1],[1.9,2],[2.9,3]$.
A big core of negative vorticity detaches from the wall and moves toward the main vortex. During its motion this big core never impinges on the wall, 
and no other vortical structures detach from the wall.}
\end{center}
\end{figure}

In the literature of the recent years it was proposed to analyze the streamwise pressure
gradient $\partial_{x}p_{w}=-\frac{1}{Re}\partial_y \omega|_{y=0}$ at the wall in order to evaluate when the viscous--inviscid
interaction begins: this was performed in \cite{Cas00,OC02}  for the thick core vortex case, and in
\cite{GSS11} for the rectilinear vortex. The idea behind this approach
lies in the fact that for Prandtl's equation the streamwise pressure gradient is imposed by the outer
flow and, in the case of stationary Euler solution, it is constant in time:
therefore any variations of  $\partial_{x}p_{w}$ in Navier-Stokes
solution is a good indicator of the beginning of the viscous-inviscid interaction.
In Figs.\ref{dpdx103}-\ref{dpdx105} we show the time evolution of
$\partial_{x}p_{w}$ starting at $t=0.2$ until $t=0.5$, with
increments of 0.1 for $Re=10^3,5\cdot10^4$ . It was proposed in \cite{GSS11} that the interaction 
begins when an inflection point in $\partial_{x}p_{w}$ forms close to its maximum, and in this case this happens
at time $t_{LS}\approx0.33,0.34,0.38$ for $Re=10^3,10^4,5\cdot10^4$ respectively.
The time of this first interaction, which in the literature is known as large--scale (LS) interaction scale , is quite early
with respect to theoretical prediction of boundary layer theory,
according to which the  viscous--inviscid interaction
begins at the singularity time. Moreover \textit{LS}-interaction has no resemblance with the
viscous-inviscid interaction developed by Prandtl's solution at $t_s$: in fact no vorticity ejection from the boundary layer is visible, 
and no large gradient in the streamwise 
variable forms. For instance in Fig.\ref{vor103_t05} the vorticity is shown at $t=0.5$ along with the pathlines of same particles
fluid followed from $t=0.4$ to $t=0.5$: no vorticity ejection
phenomena are visible (they appear at later time) and no spiky behavior is visible in the vorticity as it happens in Prandtl's vorticity at $t_s$
(see Fig.\ref{prasing}). \\
\subsection{SS interaction}
\begin{figure}
\begin{center}
\subfigure[$Re=10^4,t=0.5$]{\hskip-0.0cm\includegraphics[width=6cm]{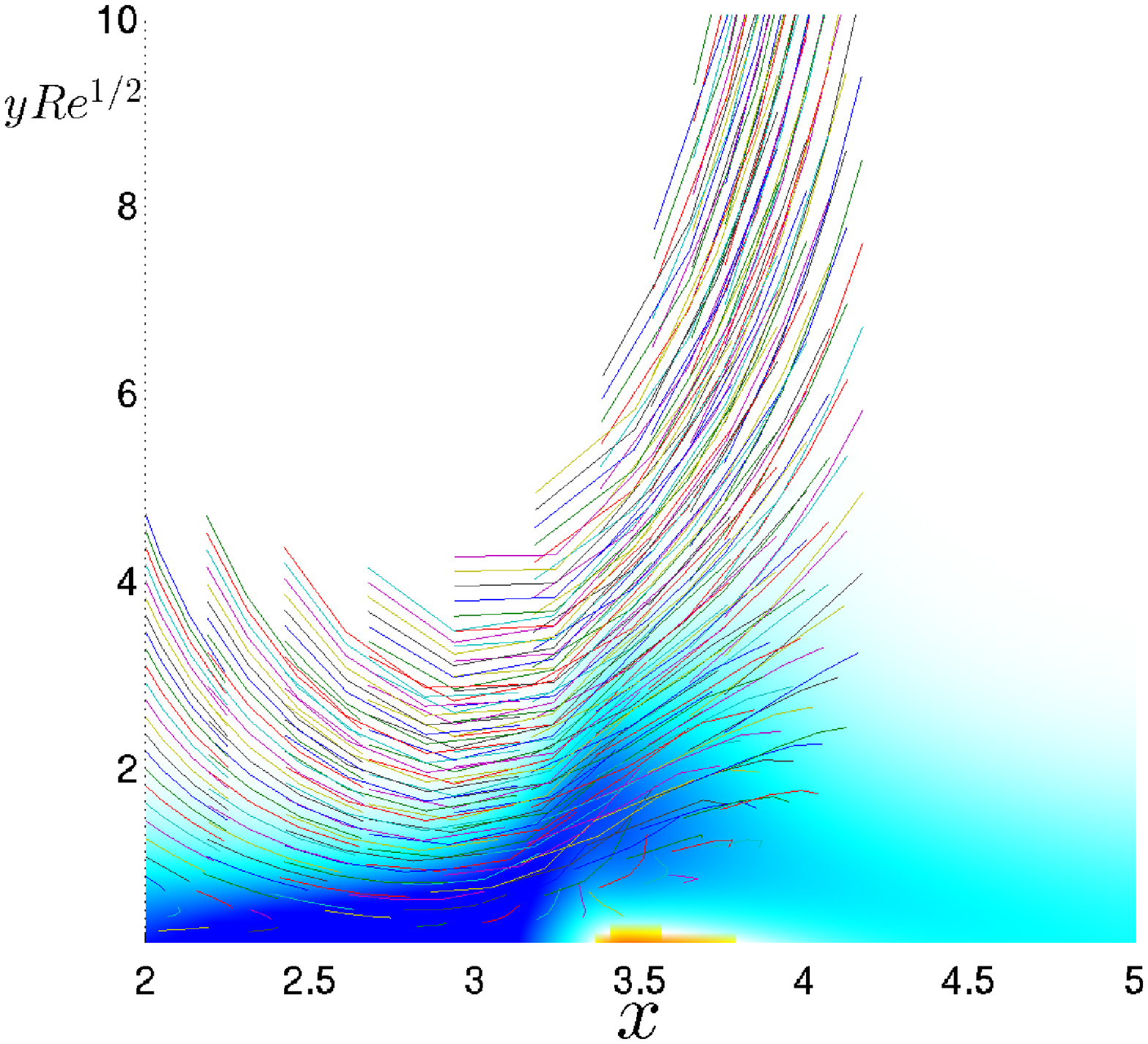}\label{vor104_t05}}
\subfigure[$Re=10^4,t=1$]{\hskip-0.3cm\includegraphics[width=6cm]{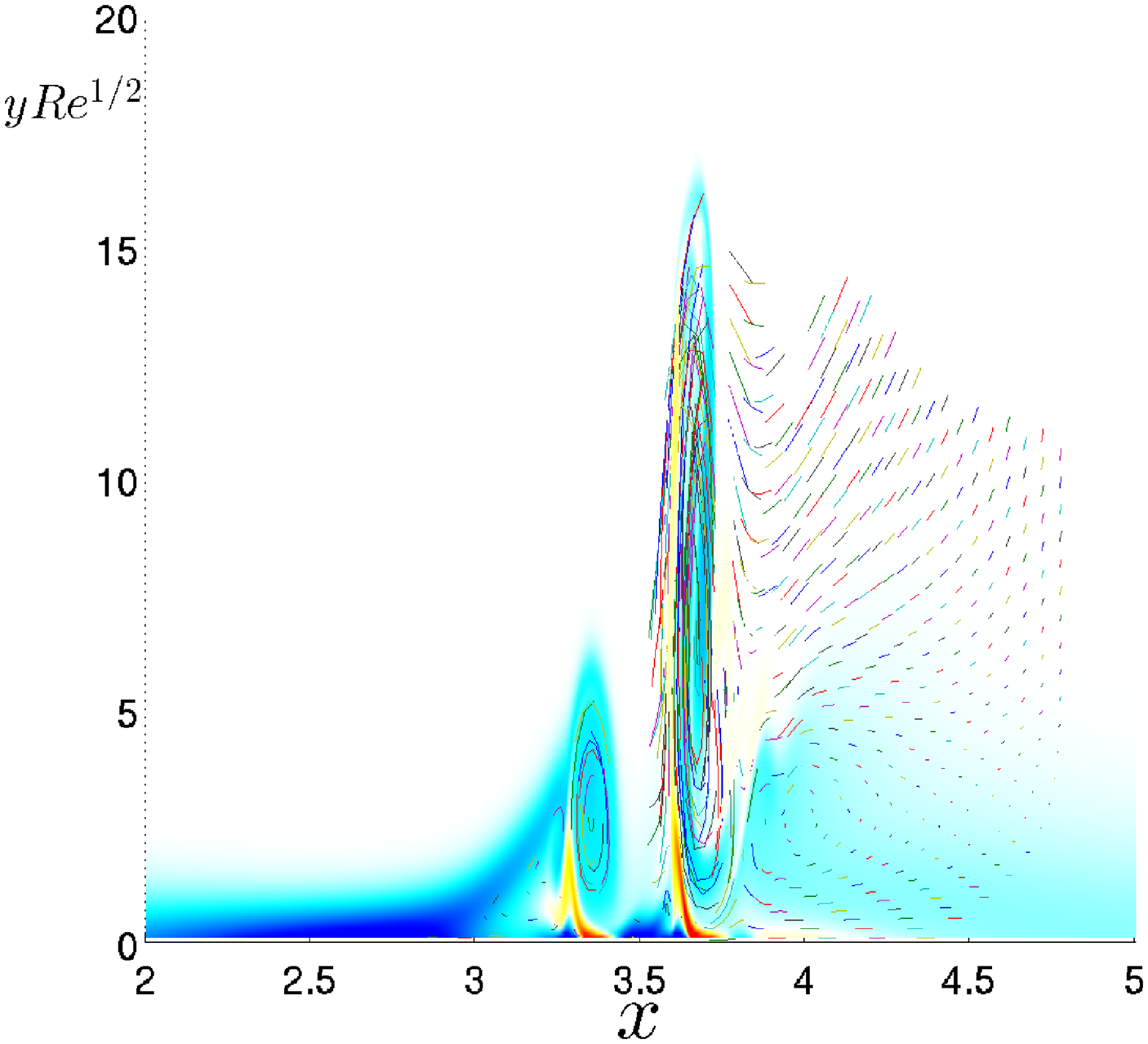}\label{vor104_t1}}
\subfigure[$Re=10^4,t=2$]{\hskip-0.0cm\includegraphics[width=6cm]{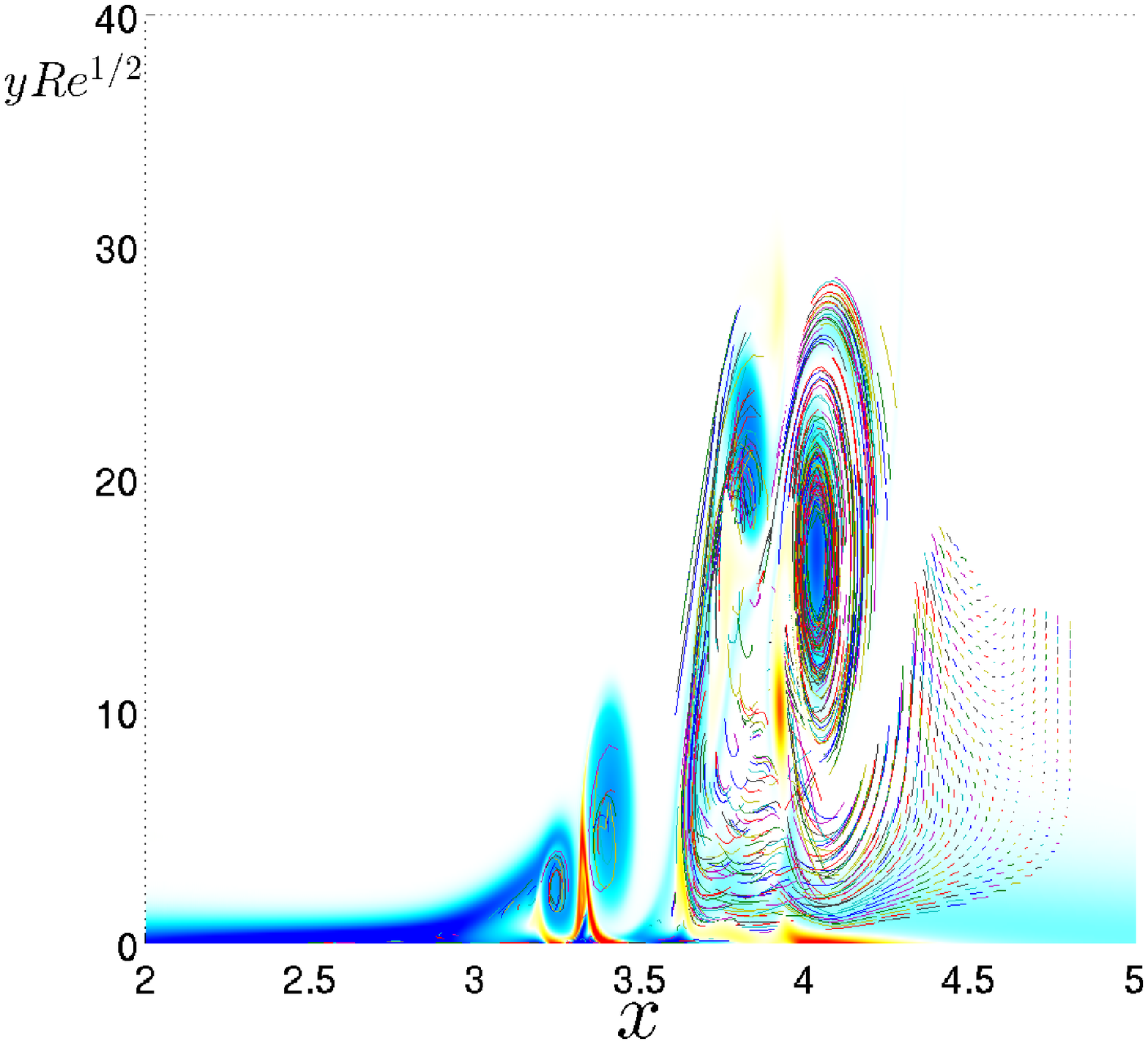}\label{vor104_t15}}
\subfigure[$Re=10^4,t=3$]{\hskip-0.3cm\includegraphics[width=6cm]{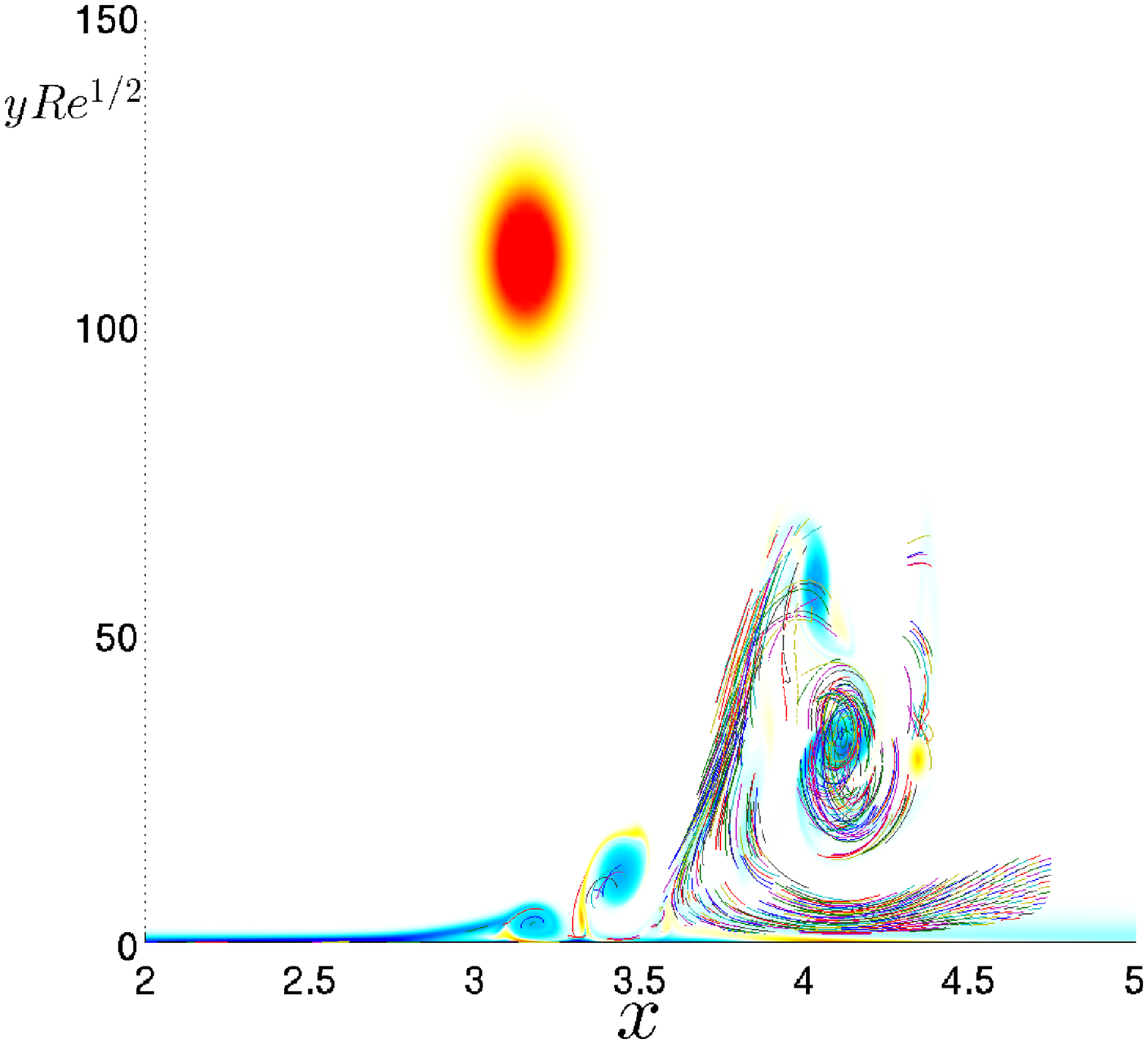}\label{vor104_t2}}
\caption{ The vorticity (red colors positive vorticity, blue color negative vorticity) and the pathlines of fluid particles  for $Re=10^4$
at time $t=0.5,1,1.5,2$. The pathlines are followed in the temporal intervals $[0.4,0.5],[0.9,1],[1.4,1.5],[1.9,2]$.
Several cores of negative vorticity detach from the wall and moves toward the main vortex. During their motion these structures impinge on the wall
leading to the growth of the enstrophy (see Fig.\ref{enstrofia}).}
\end{center}
\end{figure}

\begin{figure}
\begin{center}
\includegraphics[width=8cm]{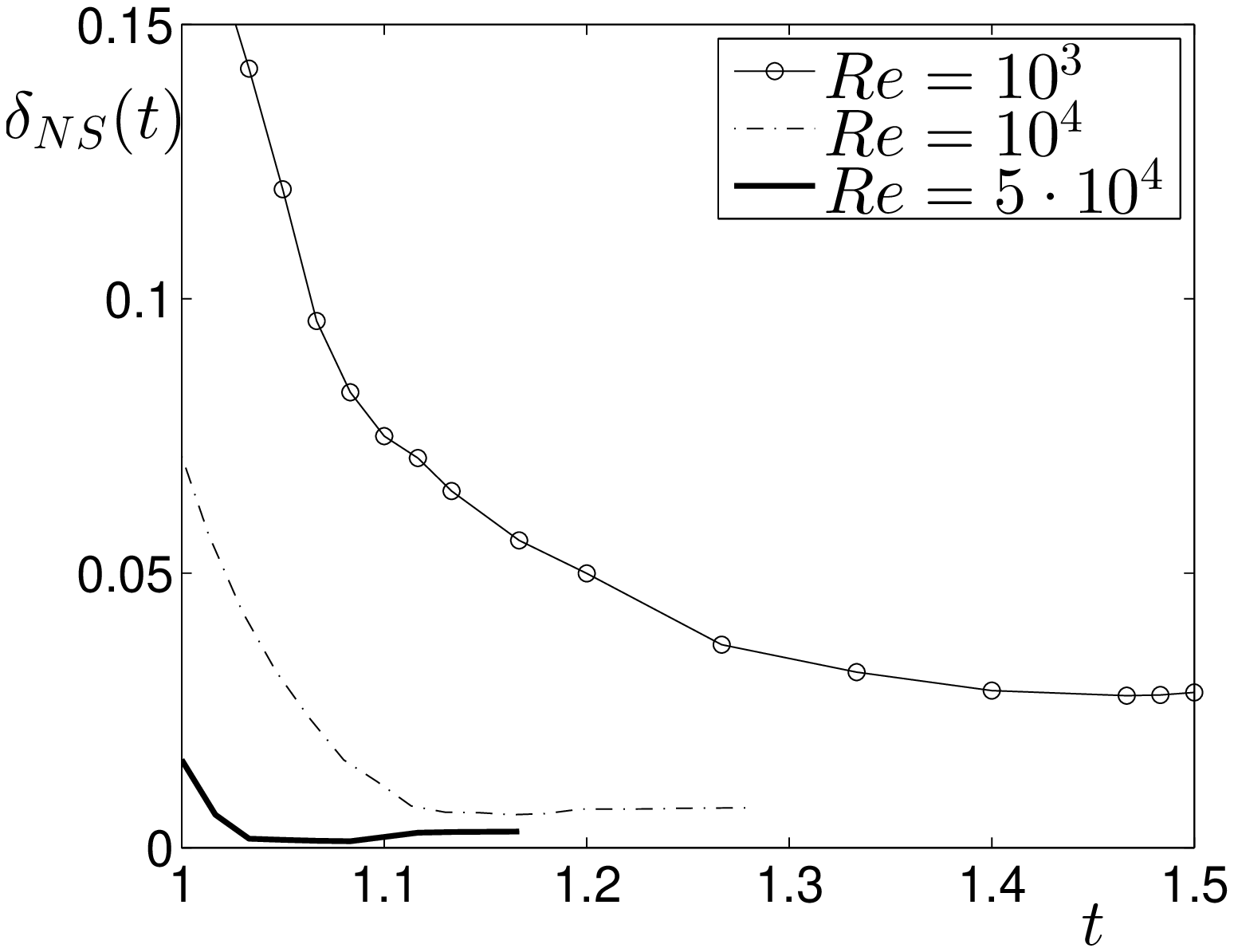}\label{delta_NS}
\caption{ Temporal evolution of $\delta_{NS}$ for various Reynolds
number.
At time $t_s=1$, $\delta_{NS}$ approaches zero as Reynolds number increases. Then it
reaches a minimum ($\delta_{NS}^m$) . The time at which 
$\delta_{NS}^m$ forms becomes closer to $t_s$ as Reynolds number increases.}
\end{center}
\end{figure}

The $LS$-interaction is the precursor, in the case of moderate--high $Re$ numbers ($Re\geq O(10^4)$),
of another interaction  acting on a smaller scale ($SS$-interaction).
This new interaction is characterized by the formation of large gradients in the streamwise variable and, physically, by the continuous formation of 
small-scale vortical structures within the boundary layer.  These structures first separate from the wall; then, due to their reciprocal interactions, they are driven
back toward the wall,
leading to a large vorticity production and the consequent growth of the enstrophy and decrease of energy.
To show how the evolution of the enstrophy is a good indicator of this phenomenology, we write the equation governing the enstrophy evolution
in the boundary layer $D=[0,2\pi]\times[0,Y_{BL}]$ where $Y_{BL}$ is large enough so that all the relevant phenomena occurring in the boundary layer are captured.
This equation is:
\begin{eqnarray}
\frac{d\Omega(t)}{dt}&=&-\frac{2}{Re}\lVert\mathbf{\nabla}\omega\rVert_{L^2(D)}^2+
2I^p(t)+NT \label{enstrophy}
\end{eqnarray}
where $\Omega(t) = \lVert \omega \rVert_{L^2(D)}^2$ and:
\begin{eqnarray}
I^p(t)= \int_{0}^{2\pi}\omega_{|y=0}\cdot\partial_x p_w
dx,\quad NT=\frac{2}{Re} \int_{0}^{2\pi}\left(\omega\cdot \partial_y\omega\right)_{|y=Y_{BL}}
dx \; . \nonumber
\end{eqnarray}
The $NT$ term is negligible because at $y=Y_{BL}$ the vorticity $\omega$ is very small, and therefore
we shall not consider this terms in the rest of our analysis.
The only way for the enstrophy
to increase is via the integral term $I^p$
which is related to the vorticity and to the vorticity flux at the boundary:  when a vortical structure impinges
on the wall a large amount of vorticity is produced and the enstrophy therefore grows.
In Fig.\ref{enstrophy} it is shown the enstrophy evolution for Prandtl's solution and for the various NS solutions (in this case we rescale the enstrophy
with a factor $Re^{-1/2}$). While for Prandtl and  NS at $Re=10^3$ the enstrophy decreases monotonically, several peaks are present for $Re=10^4,5\cdot10^4$.
For these high $Re$ numbers the vortical structures form continuously  and after the
detachment from the boundary they impinge on the wall. This leads to 
a strong ejection of flow particles from the boundary  and the growth of the enstrophy ( see  Figs.\ref{vor104_t05}-(d) for the pathlines and the vorticity contour
at $Re=10^4$).  On the other hand, the behavior at lower $Re$ numbers is quite different. In fact for $Re=10^3$ the enstrophy decreases because the
big core of negative vorticity 
that form during the separation process,  visible in Figs.\ref{vor103_t05}-\ref{vor103_t3} at different times,
 detaches from the wall,  moves toward the main vortex until they interact, and does not impinge on the wall.
This different temporal behavior of the enstrophy for the two $Re$ number regimes was also observed for the rectilinear vortex case \cite{GSS11}, and 
for a dipole-vortex \cite{KCH07,OR90}. We observe that the first peak in the enstrophy evolution forms
earlier  for $Re=5\cdot10^4$ than for $Re=10^4$, although the $LS$-interaction forms earlier for $Re=10^4$.
This confirms what observed in \cite{Cas00,GSS11} for other flow, i.e. that the
$LS$-interaction formation strongly accelerates the $SS$-interaction because the temporal gap between the beginning of the two 
interaction decreases as $Re$ increases. \\
The physical effects of the $SS$-interaction have similarities with the effect due to the singularity formation for Prandtl's equation: in fact,
there are large gradients forming in the solution  in the streamwise variable, and a strong eruption of flow particles from the boundary layer.
This leads to the conjecture that as $Re\rightarrow\infty$, $LS$ and $SS$ interactions merge together and they are a unique interaction forming at the time at which
Prandtl's singularity occurs. To give strength to this conjecture we can perform the singularity analysis to NS solution,
and in particular
we  analyze the streamwise velocity component $u$ of NS solution within the boundary region $D$ defined at the beginning of this section.
We track the
the width of analyticity strip $\delta_{NS}$
and the characterization $\alpha_{NS}$ of the main complex singularity 
of the solution by applying the singularity tracking method. In Fig.\ref{delta_NS} we can observe that  $\delta_{NS}(t)$
has a minimum $\delta_{NS}^m$ that forms closer to $t_s$ as $Re$ increases.
We can therefore expect that as $Re\rightarrow\infty$ the main complex singularity of NS solution behaves like Prandtl's singularity
and $\delta_{NS}^m\rightarrow0$ as $Re\rightarrow\infty$.
Regarding the characterization $\alpha_{NS}$ we find  for all the $Re$ that $\alpha_{NS}\approx1/2$, while $\alpha_{P}\approx1/3$.
A similar characterization was also detected in \cite{GSSC14} for impulsively started disk initial datum. In this work it was supposed 
that $\alpha_{NS}$ and $\alpha_{P}$ are strongly influenced by the viscous-inviscid interactions occurring during 
 the flow evolution. Therefore a discrepancy between the values $\alpha_{NS}$ and $\alpha_{P}$ is likely to occur.
 
\begin{figure}
\centering
\includegraphics[width=7.5cm]{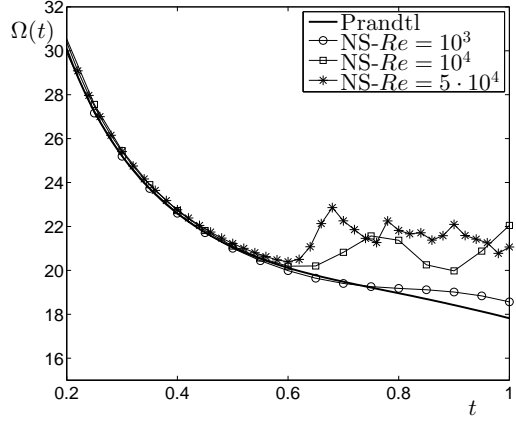}
\caption{The temporal evolution of Prandtl enstrophy $\Omega_P(t)=\lVert \partial_Y u\rVert_{L^2}^2$ and the rescaled NS enstrophy 
$\Omega(t)/\sqrt{Re}$ at different $Re$ number. Up to $LS$-interaction
the good comparison reflects the good agreement between NS and Prandtl's solutions.
During the $SS$-interaction the enstrophy for $Re=10^4-5\cdot10^4$ 
strongly differ from  the cases $Re=10^3$  and Prandtl: this is due to the interactions of the vortical structures within the boundary layer.}
\label{enstrofia}
\end{figure}

\section{Conclusion}
We have computed the solutions of 2D Prandtl and Navier-Stokes equations in the the case of a vortex array interacting with a wall.
Prandtl's equations develop a  separation singularity at  $t_s\approx1$.
The unsteady separation process for Navier-Stokes solutions reveals a different behavior from Prandtl's case. 
In particular there are two kinds of interaction between the viscous boundary layer and the inviscid outer flow .
The $LS$ interactions is found to occur for all Reynolds numbers considered, and it
is characterized by discrepancies arising between the streamwise pressure gradient at the wall and the same quantity imposed by Prandtl's solution.
The $SS$ interaction
occurs only for moderate to high Reynolds numbers ($Re\geq O(10^4$)), and it is marked by the
formation of large gradients in the streamwise variable,
and the formation of vortical structures within the boundary layer leading to a complicate flow dynamics revealed also by  the growth of the enstrophy. 
This growth, caused by the collision of the vortical structures on the wall, is absent both in Prandtl solutions as well in NS solutions for low $Re$ number.\\
The various interactions occurring in Prandtl and Navier-Stokes solutions have been
investigated by performing a complex singularity analysis on the streamwise velocity component
$u$. This analysis shows that Prandtl's singularity can be characterized as a cubic-root singularity.
On the other hand
the width  of the analyticity strip of the
Navier-Stokes solutions reaches a minimum value $\delta_{NS}^m$ decreasing as Reynolds increases, and it forms in a time which is closer to $t_s$ as Reynolds number
increases,
supporting the conjecture that $\delta_{NS}^m\rightarrow0$ at $t_s$ as
$Re\rightarrow\infty$. The
primary difference between the analysis of the spectrum of $u$ as compared to that from Prandtl's solution is the characterization obtained from the
rate of algebraic decay of the shell summed amplitudes.  For Navier-Stokes solutions 
it has been found that $\alpha^{NS}\approx1/2$ for all Reynolds numbers considered, while
the prediction of boundary-layer theory is that $\alpha^P\approx1/3$.  This discrepancy 
can be explained by the presence of the $LS$  and $SS$ interactions
that act in a different manner on the flow evolution as compared to the viscous-inviscid 
interaction present as $Re\rightarrow\infty$. We plan to investigate on the characterization of Navier-Stokes main complex singularity
for very high $Re$ numbers ($Re\geq10^6$). Moreover these very high $Re$ are characterized by the possible presence of a Rayleigh instability (observed in \cite{CO10} for
the thick-core vortex case): it is our intention to detect  if this instability is present also for other initial data  and if there is any link with the complex singularity
of Navier-Stokes solution.

\bibliographystyle{plain}
\bibliography{prandtl}

\begin{thebibliography}{10}

\bibitem{BC11}
H.~{Beir\~{a}o da Veiga} and F.~Crispo.
\newblock {C}oncerning the ${W}^{k,p}$-inviscid limit for 3-{D} flows under a
  slip boundary condition.
\newblock {\em J. Math. Fluid. Mech.}, 13:117--135, 2011.

\bibitem{BC12}
H.~{Beir\~{a}o da Veiga} and F.~Crispo.
\newblock The {3D} inviscid limit result under slip boundary conditions. a
  negative answer.
\newblock {\em J. Math. Fluid. Mech.}, 14:55--59, 2012.

\bibitem{CS97}
R.~Caflisch and M.~Sammartino.
\newblock {Navier-Stokes equations on an exterior circular domain: construction
  of the solution and the zero viscosity limit}.
\newblock {\em Comptes Rendus de l'Académie des Sciences - Series I -
  Mathematics}, 324(8):861 -- 866, 1997.

\bibitem{CLS01}
M.~Cannone, M.C. Lombardo, and M.~Sammartino.
\newblock {Existence and uniqueness for the {P}randtl equations}.
\newblock {\em Comptes Rendus de l'Académie des Sciences - Series I -
  Mathematics}, 332(3):277 -- 282, 2001.

\bibitem{CLS13}
M.~Cannone, M.C. Lombardo, and M.~Sammartino.
\newblock {Well-posedness of {P}randtl equations with non-compatible data}.
\newblock {\em Nonlinearity}, 26(3):3077--3100, 2013.

\bibitem{Cas00}
K.W. Cassel.
\newblock A comparison of {Navier-Stokes solution}s with the theoretical
  description of unsteady separation.
\newblock {\em Phil. Trans. R. Soc. Lond. A.}, 358:3207--3227, 2000.

\bibitem{CO10}
K.W. Cassel and A.V. Obabko.
\newblock A {Rayleigh} instability in a vortex-induced unsteady boundary layer.
\newblock {\em Physica Scripta}, 2010(T142):014006, 2010.

\bibitem{CMR98}
T.~Clopeau, A.~Mikelic, and R.~Robert.
\newblock On the vanishing viscosity limit for the 2d incompressible
  navier-stokes equations with the friction type boundary conditions.
\newblock {\em Nonlinearity}, 11(6):1625--1636, 1998.

\bibitem{CGS12}
G.M. Coclite, F.~Gargano, and V.~Sciacca.
\newblock Analytic solutions and singularity formation for the peakon b-family
  equations.
\newblock {\em Acta Appl. Math.}, 122:419--434, 2012.

\bibitem{CKV14}
P.~Constantin, I.~Kukavica, and V.~Vicol.
\newblock {On the inviscid limit of the {N}avier-{S}tokes equations}.
\newblock {\em arXiv:1403.5748v1}, 2014.

\bibitem{DLSS06}
G.~{Della Rocca}, M.C. Lombardo, M.~Sammartino, and V.~Sciacca.
\newblock Singularity tracking for {C}amassa-{H}olm and {P}randtl's equations.
\newblock {\em Appl. Numer. Math.}, 56(8):1108--1122, 2006.

\bibitem{GSS09}
F.~Gargano, M.~Sammartino, and V.~Sciacca.
\newblock Singularity formation for {Prandtl's equations}.
\newblock {\em Physica D: Nonlinear Phenomena}, 238(19):1975--1991, 2009.

\bibitem{GSS11}
F.~Gargano, M.~Sammartino, and V.~Sciacca.
\newblock {High Reynolds number Navier-Stokes solutions and boundary layer
  separation induced by a rectilinear vortex}.
\newblock {\em Computers \& Fluids}, 52:73--91, 2011.

\bibitem{GSSC14}
F.~Gargano, M.~Sammartino, V.~Sciacca, and K.W. Cassel.
\newblock {A}nalysis of complex singularities in high-{R}eynolds-number
  {N}avier-{S}tokes solutions.
\newblock {\em In press on J. Fluid. Mech., doi:10.1017/jfm.2014.153,
  arXiv:1310.3943.}

\bibitem{GVD10}
D.~Gerard-Varet and E.~Dormy.
\newblock On the ill-posedness of the {P}randtl equation.
\newblock {\em J. Am. Math. Soc.}, 23:591--609, 2010.

\bibitem{GVN12}
D.~Gerard-Varet and T.~Nguyen.
\newblock Remarks on the ill-posedness of the {P}randtl equation.
\newblock {\em Asymptot. anal.}, 77:71--88, 2012.

\bibitem{IP06}
D.~Iftimie and G.~Planas.
\newblock {Inviscid limits for the Navier-Stokes equations with Navier friction
  boundary conditions}.
\newblock {\em Nonlinearity}, 19:899--918, 2006.

\bibitem{Kato84}
T.~Kato.
\newblock {\em Remarks on the Zero Viscosity Limit for Nonstationary
  {N}avier-{S}tokes Flows with Boundary}, volume~2.
\newblock Springer, New York, 1984.

\bibitem{Kel06}
J.P. Kelliher.
\newblock {Navier-Stokes equations with Navier boundary conditions for bounded
  domain in the plane}.
\newblock {\em J. Math. Anals.}, 38:210--232, 2006.

\bibitem{Kel07}
J.P. Kelliher.
\newblock {On {K}ato's conditions for vanishing viscosity.}
\newblock {\em Indiana Univ. Math. J.}, 56(4):1711--1721, 2007.

\bibitem{KCH07}
W.~Kramer, H.J.H. Clercx, , and G.J.F. van Heijst.
\newblock Vorticity dynamics of a dipole colliding with a no-slip wall.
\newblock {\em Physics of Fluids}, 19(12):126603, 2007.

\bibitem{KV13}
I.~Kukavica and V.~Vicol.
\newblock {On the local existence of analytic solutions to the {P}randtl
  boundary layer equations.}
\newblock {\em Commun. Math. Sci.}, 11:269--292, 2013.

\bibitem{Lamb}
H.~Lamb.
\newblock {\em Hydrodynamics}.
\newblock Cambridge Mathematical Library. Cambridge University Press,
  Cambridge, sixth edition, 1993.
\newblock With a foreword by R. A. Caflisch [Russel E. Caflisch].

\bibitem{LCS01}
M.C. Lombardo, R.E. Caflisch, and M.~Sammartino.
\newblock {Asymptotic analysis of the linearized Navier-Stokes equation on an
  exterior circular domain: Explicit solution and the zero viscosity limit}.
\newblock {\em Communications in Partial Differential Equations},
  26(1-2):335--354, 2001.

\bibitem{LCS03}
M.C. Lombardo, M.~Cannone, and M.~Sammartino.
\newblock Well-posedness of the boundary layer equations.
\newblock {\em SIAM J. Math. Anal.}, 35(4):987--1004 (electronic), 2003.

\bibitem{LML08}
M.C. Lopes~Filho, A.L. Mazzucato, and H.J. Nussenzveig~Lopes.
\newblock Vanishing viscosity limit for incompressible flow inside a rotating
  circle.
\newblock {\em Physica D: Nonlinear Phenomena}, 237(10-12):1324--1333, 2008.

\bibitem{LMLT08}
M.C. Lopes~Filho, A.L. Mazzucato, H.J. Nussenzveig~Lopes, and M.~Taylor.
\newblock Vanishing viscosity limits and boundary layers for circularly
  symmetric {2D} flows.
\newblock {\em Bull. Braz. Math. Soc.}, 39:471--513, 2008.

\bibitem{LLP05}
M.C Lopes~Filho, H.~Nussenzveig~Lopes, and G.~Planas.
\newblock On the inviscid limit for two-dimensional incompressible flow with
  navier friction condition.
\newblock {\em SIAM Journal on Mathematical Analysis}, 36(4):1130--1141, 2005.

\bibitem{Mae13}
Y.~Maekawa.
\newblock Solution formula for the vorticity equations in the half plane with
  application to high vorticity creation at zero viscosity limit.
\newblock {\em Adv. Diff. Eqns}, 18:101--146, 2013.

\bibitem{MS12}
N.Masmoudi and F.Rousset.
\newblock Uniform regularity for the {N}avier-{S}tokes equation with {N}avier
  boundary condition.
\newblock {\em Arch. Ration. Mech. Anal.}, 203:529--75, 2012.

\bibitem{OC02}
A.V. Obabko and K.W. Cassel.
\newblock {Navier-Stokes solutions of unsteady separation induced by a vortex}.
\newblock {\em J. Fluid Mech.}, 465:99--130, 2002.

\bibitem{OR90}
P.~Orlandi.
\newblock Vortex dipole rebound from a wall.
\newblock {\em Physics of Fluids A: Fluid Dynamics}, 2(8):1429--1436, 1990.

\bibitem{PMFB06}
W.~Pauls, T.~Matsumoto, U.~Frisch, and J.~Bec.
\newblock {Nature of Complex Singularities for the 2D Euler Equation}.
\newblock {\em Physica D}, 219(1):40--59, 2006.

\bibitem{PSW91a}
V.J. Peridier, F.T. Smith, and J.D.A Walker.
\newblock Vortex-induced boundary-layer separation. {Part} 1. {The unsteady
  limit problem $Re\rightarrow\infty$}.
\newblock {\em J. Fluid Mech.}, 232:99--131, 1991.

\bibitem{Pey}
R.~Peyret.
\newblock {\em Spectral Methods for Incompressible Viscous Flow}.
\newblock Springer-Verlag, New York, 2002.

\bibitem{SC98a}
M.~Sammartino and R.E. Caflisch.
\newblock {Zero viscosity limit for analytic solutions, of the
  {N}avier-{S}tokes equation on a half-space. {I}. {E}xistence for {E}uler and
  {P}randtl equations}.
\newblock {\em Comm. Math. Phys.}, 192(2):433--461, 1998.

\bibitem{SC98b}
M.~Sammartino and R.E. Caflisch.
\newblock Zero viscosity limit for analytic solutions of the {N}avier-{S}tokes
  equation on a half-space. {II}. {C}onstruction of the {N}avier-{S}tokes
  solution.
\newblock {\em Comm. Math. Phys.}, 192(2):463--491, 1998.

\bibitem{Sh92}
M.J. Shelley.
\newblock {A study of singularity formation in vortex--sheet motion by a
  spectrally accurate vortex method}.
\newblock {\em J. Fluid. Mech.}, 244:493--526, 1992.

\bibitem{SSF83}
C.~Sulem, P.L. Sulem, , and H.~Frisch.
\newblock Tracing complex singularities with spectral methods.
\newblock {\em J. Comput. Phys.}, 50:138--161, 1983.

\bibitem{TW97}
R.~Temam and X.~Wang.
\newblock {The convergence of the solutions of the Navier-Stokes equations to
  that of the Euler equations}.
\newblock {\em App. Math. Lett.}, 10:29--33, 1997.

\bibitem{vDS80}
L.L. {van Dommelen} and S.F. Shen.
\newblock The spontaneous generation of the singularity in a separating laminar
  boundary layer.
\newblock {\em J. Comp. Phys.}, 38:125--140, 1980.

\bibitem{Wei03}
J.A.C. Weideman.
\newblock {Computing the Dynamics of Complex Singularities of Nonlinear PDEs}.
\newblock {\em J. Appl. Dynamical System}, 2(2):171--186, 2003.

\bibitem{WXZ13}
L.~Whang, Z.~Xin, and A.~Zang.
\newblock Vanishing viscous limits for {3D} {N}avier-{S}tokes equations with a
  {N}avier-slip boundary condition.
\newblock {\em J. Math. Fluid. Mech.}, 14:791--825, 2012.

\bibitem{XZ04}
Z.~Xin and L.~Zhang.
\newblock On the global existence of solutions to the {P}randtl's system.
\newblock {\em Adv. Math.}, 181(1):88--133, 2004.

\end{thebibliography}

\end{document}